\documentclass[useAMS,usenatbib,fleqn]{mn2e}

\usepackage{verbatim}
\usepackage{graphicx}
\usepackage{amssymb}
\usepackage{amsmath}
\usepackage[figuresright]{rotating}

\title[Joining the Hubble Flow]{Joining the Hubble Flow: Implications for Expanding Space}
\author[Barnes, Francis, James and Lewis]{Luke A. Barnes\thanks{E-mail: luke, mfrancis, jbjames, gfl@physics.usyd.edu.au}, Matthew J. Francis, J. Berian James and Geraint F. Lewis
\\
School of Physics, University of Sydney, NSW, Australia}

\begin{document}

\date{not yet submitted}

\pagerange{\pageref{firstpage}--\pageref{lastpage}} \pubyear{2006}

\maketitle

\label{firstpage}

\begin{abstract}

The concept of expanding space has come under fire recently as being inadequate and even misleading in describing the motion of test particles in the universe. Previous investigations have suffered from a number of shortcomings, which we seek to correct. We study the motion of test particles in the universe in detail, solving the geodesic equations of General Relativity for a number of cosmological models. In particular, we use analytic methods to examine whether particles removed from the Hubble flow asymptotically rejoin the Hubble flow, a topic that has caused confusion because of differing definitions and invalid reasoning. We conclude that particles in eternally expanding but otherwise arbitrary universes do not in general rejoin the Hubble flow.

\end{abstract}



\section{Introduction}
The paradigm of modern cosmology is the Friedmann-Lema\^\i tre-Robertson-Walker (FLRW) model \citep{friedmann,1931MNRAS..91..483L,1935ApJ....82..284R,walker}. In recent years there has been debate over paradoxical features of the model, and the physical interpretation of its dynamics. Attention has been drawn in particular to superluminal recession velocities \citep{2001AIPC..555..348D,2001astro.ph..4349D,2006astro.ph..1171C,2006astro.ph..2102S} and to the motion of test particles in expanding space \citep{2001astro.ph..4349D,2004Obs...124..174W,2006astro.ph..3162G,Peacockweb}. It is to the second of these issues that we turn our attention.

The motion of test particles in the FLRW model is a fascinating illustration of the interaction between physical concepts and quantitative theories. One of the defining characteristics of physics is the mathematical precision of its predictions. Yet there is more to applying physical laws than simply solving equations. In order to make physical laws more transparent and accessible, we use physical concepts that develop an intuition or a mental picture of the scenario. A successful physical concept allows us to shortcut the mathematics, qualitatively understanding a scenario without having to solve the equations. As an example, consider the north pole of a bar magnet approaching a loop of wire. Looking from behind the magnet, we know that an anti-clockwise current will be set up in the wire, but we need not come to this conclusion by solving Maxwell's equations---Lenz's law will give us the answer without the mathematics.

Attached to the equations of the FLRW model is the physical concept that ``space is expanding''. Galaxies, we are taught, are receding not because they are moving through space but because space itself is being stretched between us and the galaxy. On the face of it, this concept gives us a good intuitive understanding of many cosmological phenomena---it helps us understand why the velocity-distance law is linear and why light is redshifted as it moves through the universe. However, it has been attacked recently, most notably by \citet{2004Obs...124..174W} and \citet{Peacockweb}, as being inadequate in describing local dynamics. Whilst \citet{Peacockweb} will allow a global form of the expanding space concept---the total volume of a closed universe increases with time---he contends that:
\begin{quote}
there is no local effect on particle dynamics from the global expansion of the universe \ldots `Expanding space' is in general a dangerously flawed way of thinking about an expanding universe.
\end{quote}

Previous attempts at resolving this debate have suffered from a number of shortcomings. The most common is the overwhelming desire to approximate General Relativity (GR) by something else---Newtonian gravity, Special Relativity (SR) or a weak field limit of GR. This is probably wrong and certainly unnecessary---why approximate when you can use the exact geodesic equations? Another problem is the small range of cosmological models considered---\citet{2006astro.ph..3162G}, for example, do a marvellous job of solving the geodesic equations and then apply the solution to just two models, both of which are observationally disfavoured. There is also a dangerous reliance on numerical calculations to determine asymptotic ($t \to \infty$) behaviour. \citet{2004Obs...124..174W} makes this point about \citet{2001astro.ph..4349D}, but doesn't say why analytic solutions to a Newtonian approximation are better than numerical solutions of the exact GR equations.

In this paper we address these problems by treating the asymptotic behaviour of the full GR equations analytically, for general (up to non-oscillating, bounded equations of state) cosmological models. Section \ref{GRequations} will give the full equations of motion for particles in the FLRW model. In Section \ref{testparticles} we will analyse test particle motion and correct some errors made in previous work. In Section \ref{joininghubble} we consider the asymptotic behaviour of particles in the universe, and in particular the notion of joining the Hubble flow. Finally, in Section \ref{space} we briefly outline the relevance of our results for expanding space. Our aim is not to be the final word in this debate but instead to clarify the relevant features of Robertson-Walker spacetime.


\section{The Cosmological GR Equations of Motion}\label{GRequations}
We begin by giving the equations for the FLRW model. The Robertson-Walker (RW) metric, which tells us how to measure distance and time in a homogeneous and isotropic universe, has the line element:
\begin{multline} \label{eq:RW}
\textrm{d}s^2 = c^2 \textrm{d}t^2 - R^2(t) \left(\textrm{d}\chi^2 + S_k^2 \left(\textrm{d}\theta^2 + \sin^2\theta \textrm{d}\phi^2\right)\right)
\end{multline}
where $c$ is the speed of light (hereafter $c = 1$, unless reintroduced for clarity) and $S_k(\chi) = \sin\chi, \chi, \sinh\chi$ for $k = +1,0,-1$, where $k$ is the curvature constant for closed, flat and open space respectively. The RW metric relates the spacetime interval $\textrm{d}s$ to the cosmic time $t$ and the spherical comoving coordinates $(\chi,\theta,\phi)$. More will be said about these coordinates later. The scale factor $R(t)$ is the key prediction of any cosmological model, encapsulating the beginning, evolution and fate of the universe. We can also define the Hubble parameter $H$, which measures the rate of expansion of the universe:
\begin{equation} \label{eq:Hdef}
H \equiv \frac{1} {R} \frac{\textrm{d}R} {\textrm{d}t}
\end{equation}

In the curved, expanding spacetime of the RW metric, we must be very careful when defining distance measures (see \citet{Linder:1997} for details and \citet{Hogg:1999ad} for a summary). Throughout this paper, we will use proper distance $r_\textrm{p}$, which is defined as being the radial ($\textrm{d}\theta = \textrm{d}\phi = 0$) spacetime interval ($\textrm{d}s$) along a hypersurface of constant cosmic time ($\textrm{d}t = 0$)\footnote{A thought experiment for measuring proper distance is as follows: we imagine being at one end of a giant ruler, pointed at a distant object. A volunteer is sent along the ruler to read off the distance to the object. Since the universe is expanding, the volunteer will need to carry a clock that displays cosmic time, and note down the time when the measurement was made. When light rays have carried the volunteer's result back to us, we will know the proper distance to the object at the time the measurement was made. \citet{2005astro.ph.12282S} criticises proper distance as ``violating the principle that instantaneous non-local measurements cannot be made''. This amounts to criticising a spacelike interval for being a spacelike interval. In any GR metric, length or distance is defined as the spacetime interval along a surface of constant time, and as such can never be known instantaneously. This does not mean, however, that proper distance is unphysical. It only means that it must be reconstructed at a later time from the information in light signals.}. The RW metric then gives the proper distance between the origin ($\chi = 0$) and $\chi$ at time $t$ to be:
\begin{equation} \label{eq:prop}
r_\textrm{p}(t) = R(t) \chi(t)
\end{equation}

The field equations of GR allow us to find $R(t)$ given the energy content of the universe. The result is the Friedmann equations, which, following \citet{2005gere.book.....H}, equation 15.13, we will write as:
\begin{equation} \label{eq:fried}
H^2 = H_0^2 \sum_i \Omega_{i,0} \left( \frac{R} {R_0} \right)^{-3(1+w_i)}
\end{equation}
The other two Friedmann equations can be found in \citet{2005gere.book.....H}, but will not be needed here. A subscript zero always refers to a quantity evaluated at the present epoch. The sum is over the energy components of the universe (labelled $i$), each with corresponding equation of state $w_i = p_i/\rho_i$, where $p$ is the pressure and $\rho$ is the energy density. Equation \eqref{eq:fried} assumes that the energy components do not interact and that each $w_i$ is a constant\footnote{Evolving equations of state and interacting components will only be considered with regards to their asymptotic behaviour---cf. footnote \ref{evolfoot}; see \citet{2005PASA...22..315B} and references therein for details on such cosmological models.}, as it is for most familiar forms of energy---matter ($w_m = 0$), radiation ($w_r = 1/3$), vacuum energy ($w_{\Lambda} = -1$). The sum includes ``curvature energy'' ($\Omega_{k,0} = -k/(R_0H_0)^2 = 1 - \sum_{i \neq k} \Omega_{i,0}$) which has $w_k = -1/3$. This can be thought of as convenient shorthand. A particular solution to the Friedmann equations which will prove useful in Section \ref{comparing} is the case of a universe with a single component $w > -1$:
\begin{equation} \label{eq:flat}
R(t) = R_0\left( \frac{t} {t_0} \right)^{\frac{2} {3(1+w)}}
\end{equation}
where $t_0$ is the age of the universe. For $w = -1$, we have the solution $R(t) = R_0\exp\left( \frac{t-t_0} {t_0}\right)$, where $t_0$ is the e-folding time for the expansion. We will not consider phantom energy with $w < -1$ \citep[see][and references therein for details]{2003PhRvL..91g1301C}.

The trajectory of a particle in the universe is computed by solving the geodesic equations of GR, given by:
\begin{equation} \label{eq:geodesic1}
\frac{\textrm{d}^2 x^a} {\textrm{d} \lambda^2} + \Gamma^a_{bc} \frac{\textrm{d}x^b} {\textrm{d}\lambda} \frac{\textrm{d}x^c} {\textrm{d}\lambda} = 0
\end{equation}
where $\lambda$ is an affine parameter, the indices $a,b,c$ run from $0$ to $3$\ and $(x^0, x^1, x^2, x^3) = (t,\chi,\theta,\phi)$. The Christoffel symbol $\Gamma^a_{bc}$ is given by:
\begin{equation} \label{eq:christoffel}
\Gamma^a_{bc} = \frac{1} {2} g^{ad} \left( \partial_b g_{cd} + \partial_c g_{bd} - \partial_d g_{bc} \right)
\end{equation}

For the RW metric, we have 
\begin{equation}
g_{ab} = diag(1,-R^2,-R^2 S_k^2(\chi),-R^2S_k^2(\chi) \sin^2\theta)
\end{equation}
which gives:
\begin{subequations} \label{eq:geodesicfull}
\begin{align}
t'' + R\dot{R} \left( \chi'^2 + S_k^2(\chi)\left( \theta'^2 + \sin^2\theta \phi'^2 \right) \right) =& 0 \label{eq:geo1} \\
\chi'' + 2H t'\chi' - S_k(\chi) C_k(\chi) \left( \theta'^2 + \sin^2\theta \phi'^2 \right) =& 0 \label{eq:geo2} \\
\theta'' + 2H t'\theta' + 2\frac{C_k(\chi)} {S_k(\chi)} \chi'\theta' - \sin\theta \cos\theta \phi'^2 =& 0 \label{eq:geo3} \\
\phi'' + 2H t'\phi' + 2\frac{C_k(\chi)} {S_k(\chi)} \chi'\phi' +2\cot\theta \phi'\theta' =& 0 \label{eq:geo4} \\
t'^2 -R^2 \left( \chi'^2 + S_k^2(\chi)\left( \theta'^2 + \sin^2\theta \phi'^2 \right) \right) =& \alpha \label{eq:geo5}
\end{align}
\end{subequations}
where a prime means $\textrm{d} / \textrm{d}\lambda$, and an overdot means $\textrm{d} / \textrm{d}t$. $C_k(\chi) = \textrm{d}S_k(\chi) / \textrm{d}\chi = \cos\chi, 1, \cosh\chi$ for $k = +1,0,-1$. Equation \eqref{eq:geo5} is the normalisation condition for the four-velocity; we set $\alpha = -1,0,+1$ for spacelike, null and timelike geodesics respectively. For timelike geodesics, this means that the affine parameter is the proper time measured by a clock that follows a given geodesic. We will only consider timelike geodesics.

Equations \eqref{eq:geo1}-\eqref{eq:geo4} allow a trivial solution, i.e. $\chi$, $\theta$, $\phi$ = constant for all time. Thus, there exists a family of free-falling particles that maintain their position in comoving coordinates---in fact, this is what we mean by comoving coordinates. This family of particles defines the Hubble flow.

For motion in one dimension, we can choose our coordinates to make make the motion purely radial. \citet{2006astro.ph..3162G} study this case and derive the following useful equations:
\begin{subequations} \label{eq:georadial}
\begin{align}
\dot{\chi} =& (R^2 + CR^4)^{-1/2} \label{eq:georad1} \\
\chi =& \chi_0 \pm \int \frac{\textrm{d}t} {R\sqrt{1 + CR^2}} \label{eq:georad2}
\end{align}
\end{subequations}
where
\begin{equation}
C = \frac{1}{\dot\chi_0^2R_0^4} - \frac{1}{R_0^2}
\end{equation} is strictly positive. Note that these are equations for $\chi$ as a function of $t$ not $\lambda$.


\section{Motion of Test Particles}\label{testparticles}
Following previous work on this topic, we set up the initial conditions of our test particle as follows. We place ourselves at the origin ($\chi = 0$) and the test particle at $\chi_0$. We consider the particle to initially have constant proper distance $\dot{r}_\textrm{p}(t_0) = 0$ i.e.
\begin{equation} \label{eq:initialchi}
\frac{\left( R(t) \chi \right)} {\textrm{d}t} \Bigr\rvert_{t=t_0} = 0 \quad \Rightarrow \quad \dot{R}(t_0)\chi_0 + R(t_0)\dot{\chi}(t_0) = 0
\end{equation}

The reason for this particular initial condition is quite simple. A popular way of visualising expanding space is a balloon or a large rubber sheet. Imagine yourself and a friend at rest on a large rubber sheet. We cannot directly observe spacetime, so we will do this thought experiment in the dark\footnote{We will not speculate on how you two came to be standing on a rubber sheet in the dark.}. Suppose you both observe a glowing ball moving away from you. ``The rubber sheet is being stretched,'' you say. ``No it's not,'' replies your friend, ``the sheet is still and the ball is rolling away.'' Together, you come up with an ingenious way of finding out who is right. You take another glowing ball, and drop it onto the sheet a certain distance away. If the sheet is expanding, then we expect it to carry the ball away; if the sheet is still then the recession of the first ball was due to a kinematical initial condition. Once this is removed, so is the recession.

The cosmological expansion is a bit more complicated, as we have expansion that changes with time due to the self gravitation of the energy contents of the universe. We will therefore need to consider a range of cosmological models. For reasons that will become clear in the next section, we will consider models given by Equation \eqref{eq:flat}. We will allow $\chi$ to be negative when a particle passes through the origin, rather than have to worry about a change in angular coordinates. We have chosen $t_0 = 1$ and $R_0 = 1$ for each model. We started each particle off at comoving coordinate $\chi_0 = 1/3$, which ensures that $C>0$. Physically, this ensures that we do not place the particle beyond the Hubble sphere, which would require a velocity relative to the local Hubble flow (peculiar velocity) greater than the speed of light\footnote{\citet{2006astro.ph..3162G}, equation (22) gives this condition incorrectly. The correct formula, which follows directly from their equation (21), is $\chi_0 < 3(1+w)/2$. They then claim to start at particle off at $\chi_0 = 1$ in a Milne ($w=-1/3$) universe, which contradicts the previous condition on $\chi_0$. Therefore, their Figure 1b) appears to plot a null geodesic.}.

\begin{figure*}
\centering
\includegraphics[width=\linewidth]{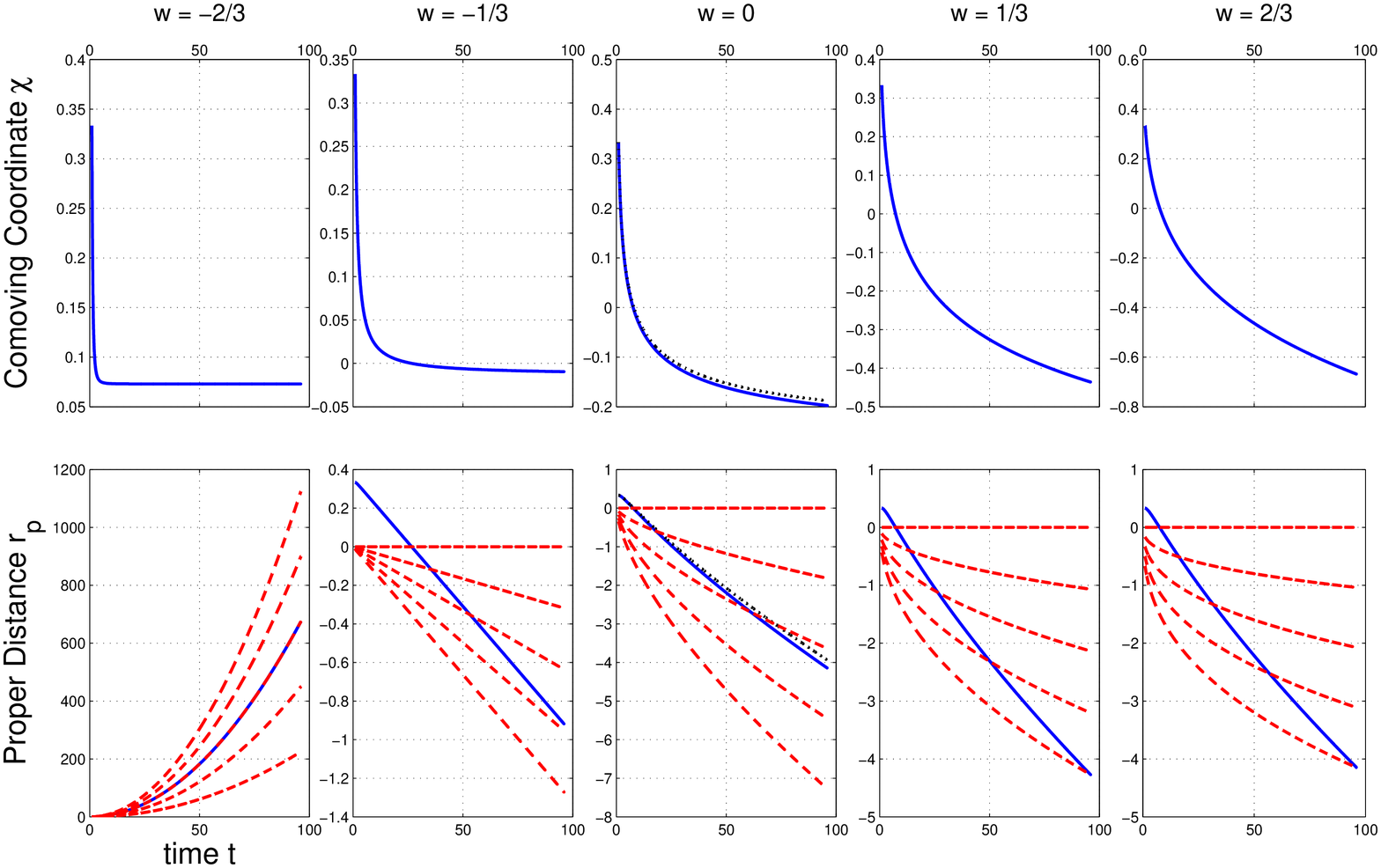}
\caption{Comoving radial coordinate and proper distance (solid lines) as a function of time for a particle in radial motion for cosmological models with differing values of the equation of state $w$ given above each column. The dashed lines in the lower panels show the motion of nearby particles in the Hubble flow. The dotted line in the centre panels gives the Newtonian solution for the motion of the particle, as discussed in the text. Note that the vertical scale changes from panel to panel.}\label{fig:chi}
\end{figure*}

\begin{figure}
\includegraphics[width=\linewidth]{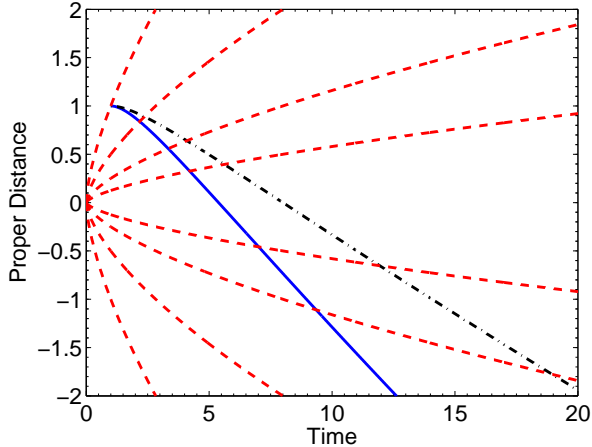}
\caption{A reproduction of Figure 1 in \citet{2004Obs...124..174W}, plotting proper distance against time for a test particle released from $\chi_0=1$ in an Einstein de-Sitter universe (i.e. flat, matter only). The dot-dashed line is the Newtonian solution, whilst the solid line is the relativistic solution. The dashed lines represent particles in the Hubble flow. The discrepancy between the solutions is obvious.}\label{fig:whiting}
\end{figure}

The results of solving the geodesic equations are shown in Figure \ref{fig:chi}. The dotted line in the centre panels shows the motion of the particle as calculated by a Newtonian analysis (see \citet{2004Obs...124..174W} and \citet{Peacockweb}). The equation of motion is:
\begin{equation}
r_\textrm{p}^{\textrm{Newton}} = 2R_0\chi_0 \left( \frac{t} {t_0} \right)^{1/3} - R_0\chi_0 \left( \frac{t} {t_0} \right)^{2/3}
\end{equation}
The Newtonian result is surprisingly accurate, remaining close to the GR solution even up to 100 times the age of the universe. It is seen to diverge from the exact solution eventually, though, and thus remains only a useful approximation for small times. This divergence becomes more apparent as we increase $\chi_0$, i.e. as we approach the Hubble sphere. If we consider the $w=0$, Einstein de-Sitter universe and set $\chi_0=1$, then the solutions diverge much more quickly. Figure \ref{fig:whiting} reproduces Figure 1 of \citet{2004Obs...124..174W}, overlaying the relativistic solution. It is easy to see that whilst the qualitative behaviour is similar, the Newtonian solution is quantitatively different.

Returning to Figure \ref{fig:chi}, a few points are noteworthy. The bottom, leftmost panel ($w = -2/3$) shows the particle trajectory moving away from the origin and very quickly becoming indistinguishable from the nearby Hubble flow. The other panels do not show the same behaviour, but instead the particle moves toward the origin and away on the opposite side of the sky. Moreover, they don't seem to be attaching themselves to any particular particle in the Hubble flow. But, as noted in the introduction, it is dangerous to try to determine asymptotic behaviour from numerical plots. We must do it analytically.


\section[Joining the Hubble Flow]{Joining the Hubble Flow}\label{joininghubble}
There is disagreement in the literature as to the fate of free particles in an eternally expanding universe, i.e. where $R(t) \to \infty$ as $t \to \infty$. \citet{2001astro.ph..4349D} and \citet{2006astro.ph..3162G} claim that they will asymptotically rejoin the Hubble flow, whilst \citet{2004Obs...124..174W} states that ``it cannot be asserted that \ldots free particles [are] swept into the Hubble flow, even asymptotically.'' This is an important issue because it is often claimed that the expansion of space will dampen out all motion through space, so that a particle initially removed from the Hubble flow will asymptotically rejoin it.

On closer inspection, this disagreement stems from different definitions of what it means to ``asymptotically rejoin the Hubble flow.'' In this section we will propose a number of precise definitions of this phrase, and then see which ones are equivalent and which ones hold in an eternally expanding but otherwise arbitrary universe. We will not consider universes in which the current expansion becomes a contraction at some point in the future. This would be an unnecessary and distracting complication when considering the expansion of space. We do not claim that this list is exhaustive.

\subsection{Seven Definitions}

\subsubsection*{\textbf{Definition 1} ($\dot{\chi}\to 0$): A particle with coordinate trajectory $\chi(t)$ asymptotically rejoins the Hubble flow if $\dot{\chi}(t) \to 0$ as $t \to \infty$.}

The particle is deemed to asymptotically rejoin the Hubble flow if its velocity through coordinate space approaches the velocity through coordinate space of the Hubble flow, namely zero.

\subsubsection*{\textbf{Definition 2 ($\chi \to \chi_{\infty}$)}: A particle with coordinate trajectory $\chi(t)$ asymptotically rejoins the Hubble flow if $\chi \to \chi_{\infty}$ as $t \to \infty$, where $\chi_{\infty}$ is a constant that depends on the cosmology and initial conditions.}

The Hubble flow is defined by having constant coordinates. Thus, we consider a particle to approach the Hubble flow if its radial coordinate approaches a constant. The asymptotic value of the radial coordinate ($\chi_{\infty}$) can be thought of as the ``rightful place'' of the particle in the Hubble flow.

\subsubsection*{\textbf{Definition 3} ($v_{\textrm{pec}} \to 0$): A particle with coordinate trajectory $\chi(t)$ asymptotically rejoins the Hubble flow if $v_{\textrm{pec}}(t) \equiv R(t)\dot{\chi}(t) \to 0$ as $t \to \infty$.}

We can divide the proper velocity ($\dot{r}_\textrm{p}$) of a test particle into a recession component and a peculiar component as follows:
\begin{equation}
\dot{r}_\textrm{p} = \dot{R}(t)\chi(t) + R(t)\dot{\chi}(t) = v_{\textrm{rec}}(t) + v_{\textrm{pec}}(t)
\end{equation}
If we move our coordinate origin so that $\chi(t) = 0$ at time $t$, then we see that the proper velocity of the test particle is solely its peculiar velocity. Thus peculiar velocity is simply proper velocity relative to the local Hubble flow. The requirement that $v_{\textrm{pec}}(t) \to 0$ as $t \to \infty$ is equivalent to the velocity relative to the local Hubble flow going to zero.

\subsubsection*{\textbf{Definition 4} ($\dot{r}_\textrm{p} \to v_{\textrm{rec}}$): A particle with coordinate trajectory $\chi(t)$ asymptotically rejoins the Hubble flow if $\dot{r}_\textrm{p} \to v_{\textrm{rec}}(t)$ as $t \to \infty$.}

We require that the proper velocity of the test particle approaches its recession velocity. This is subtly different from definition 3, as will be explained below. To avoid ambiguity, this definition uses a continuous version of asymptotic equivalence: we say that $f(x)$ approaches $g(x)$ as $x \to \infty$ if their ratio approaches unity, i.e.
\begin{equation}\label{eq:asympequiv}
f(x) \to g(x) \text{ as } x \to \infty \quad \Leftrightarrow \quad \frac{f(x)} {g(x)} \to 1 \text{ as } x \to \infty
\end{equation}

\subsubsection*{\textbf{Definition 5} ($\Delta r_{\textrm{p}}\to 0$): A particle with coordinate trajectory $\chi(t)$, where $\chi(t) \to \chi_{\infty}$ as $t \to \infty$, asymptotically rejoins the Hubble flow if $\Delta r_{\textrm{p}} \equiv \lvert R(t)\chi_{\infty} - R(t)\chi(t) \rvert \to 0$ as $t \to \infty$.}

Suppose that our test particle is approaching a particular coordinate ($\chi = \chi_{\infty}$, cf. definition 2) and that we place a reference particle in the Hubble flow at this coordinate. We require that the proper distance between the test particle and the reference particle approach zero. In other words, the test particle sees its rightful place in the Hubble flow get closer (in terms of proper distance) asymptotically. Note that if we had chosen instead to require that $R(t) \chi(t) \to R(t) \chi_{\infty}$ then this definition would have been equivalent to Definition 2 by Equation \eqref{eq:asympequiv}.

\subsubsection*{\textbf{Definition 6} ($z_{\textrm{obs}} \to z_{\textrm{cosm}}$): A particle with observed redshift $z_{\textrm{obs}}(t_r)$ at time of reception $t_r$ asymptotically rejoins the Hubble flow if $z_{\textrm{obs}}(t_r) \to z_{\textrm{cosm}}$ as $t_r \to \infty$, where $t_e$ is the time of emission of a photon that reaches the observer at $t_r$.}

Light emitted from a particle in the Hubble flow is observed to be redshifted according to the cosmological redshift formula: $z_{\textrm{cosm}}\equiv R(t_r) / R(t_e) - 1$. For a particle with coordinate trajectory $\chi(t)$, there is an additional Doppler redshift resulting from its velocity relative to the Hubble flow:
\begin{align} \label{eq:zobs}
1+z_{\textrm{obs}}(t_r) &= (1+z_{\textrm{cosm}})(1+z_{\textrm{Dop}}) \\
&= \left( \frac{R(t_r)} {R(t_e)} \right) \left( \frac{1+v_{\textrm{pec}}(t_e)} {1-v_{\textrm{pec}}(t_e)} \right)^{\frac{1} {2}}
\end{align}
where $v_{\textrm{pec}}$ is considered positive when the particle's velocity through the local Hubble flow points away from us. The Doppler term is the familiar redshift of light formula from SR, but the formula is derived purely from the RW metric.

This definition assumes that we are comoving observers that measure light signals sent from the test particle. Suppose at each time we place a reference particle in the Hubble flow at the same coordinate as the test particle. The redshift of the reference particle, which represents the local Hubble flow, will be purely cosmological. This definition requires that the redshift of the test particle approach the redshift of the reference particle.

\subsubsection*{\textbf{Definition 7} (CMB dipole $\to 0$): A particle moving through a universe containing a cosmic microwave background (CMB) asymptotically rejoins the Hubble flow if the dipole anisotropy in the CMB goes to zero as $t \to \infty$.}

In a universe filled with black-body radiation at a certain temperature $T$, an observer moving through the Hubble flow will see the CMB to be hotter in one direction and colder in the opposite direction. Indeed, this is exactly what we see from Earth---it is known as the ``great cosine in the sky'' and disrupts the isotropy of the CMB at a level of $\sim 10^{-3}$ \citep[][among others]{2003ApJS..148....1B}. The maximum difference between the temperature as measured by an observer in the Hubble flow ($T_0$) and our test particle with peculiar velocity $v_{\textrm{pec}} << c$ is given by \citep{1968PhRv..174.2168P,2002NewAR..46..693M}:
\begin{equation} \label{eq:CMBtemp}
\frac{\Delta T} {T_0} \sim \frac{v_{\textrm{pec}}} {c}.
\end{equation}

Finally, note that it is too much to ask that the test particle exactly join the Hubble flow after some time, i.e. $\chi(t) = \chi_{\textrm{f}}$, a constant, for all $t \ge t_\textrm{f}$. Only pathological functions that do not equal their own Taylor series can do this, and it is unlikely that such functions will appear as a solution to the geodesic equations \eqref{eq:geodesicfull}.

\subsection{Comparing the Definitions} \label{comparing}
The previous section may appear to be an exercise in pedantic cosmology, and one hopes that all the definitions will turn out to be equivalent. However, it turns out that only three of the above definitions hold in eternally expanding but otherwise arbitrary universes, and one of them fails in all non-accelerating universes.

We will need two key results in order to analyse these definitions. The first is that all eternally expanding cosmological models approach the single component model of Equation \eqref{eq:flat} as $t \to \infty$. We can show this directly from Equation \eqref{eq:fried}. Consider the universe to contain a number of energy components (labelled $i$), each with constant\footnote{For a component with an evolving equation of state, consider the asymptotic value of the equation of state, i.e. $w_i(t) \to w_{i,\infty}$ as $t \to \infty$. A unbounded equation of state is most likely unphysical. An oscillating equation of state will not be considered.\label{evolfoot}} equation of state $w_i$. Now consider the right hand side of Equation \eqref{eq:fried}. As $t \to \infty$, we know that $R(t) \to \infty$ since we are only considering eternally expanding universes. Since the dependence on $R(t)$ is $R^{-3(1+w_i)}$, we see that, for large $t$, the component with the most negative equation of state will dominate the dynamics of the universe. Precisely, let the component with the most negative equation of state be called the dominant component ($i = d$), with equation of state $w_d$. Then, for large $t$
\begin{align}
&H^2 \approx H_0^2 \ \Omega_{d,0} \ \left( \frac{R} {R_0} \right)^{-3(1+w_d)} \\
\Rightarrow \quad &R(t) \approx R_0\left( \frac{t} {t_0} \right)^{\frac{2} {3(1+w_d)}} \label{eq:domat}
\end{align}
which is Equation \eqref{eq:flat} with $w = w_d$. For example, in a universe with matter density less than critical, we consider ``curvature energy'' to be the dominant component with $w_d = -1/3$. The exact solution for this cosmology (given in \citet{2005gere.book.....H}, pg. 402) indeed shows that $R(t) \propto t$ for large $t$. Thus, Equation \eqref{eq:domat} is a general form for $R(t)$ when considering the asymptotic behaviour of the universe\footnote{There is a subtlety here---a universe that contains only ``curvature energy'' (i.e. an empty universe) is not identical to a universe containing a critical density fluid with equation of state $w = -1/3$. Although the dependence of the scale factor on time is the same in both universes ($R(t) \propto t$), the empty universe has $k = -1$, whilst the $w = -1/3$ fluid universe has $k = 0$. However, Equations \eqref{eq:geodesicfull} show that if we consider radial geodesics ($\theta' = \phi' = 0$), then there is no dependence on $k$. Thus, the distinction between these two universes can be ignored for now.}. We can also calculate the deceleration parameter, $q$:
\begin{equation} \label{eq:qoft}
 q \equiv - \frac{\ddot{R}} {R} \frac{1} {H^2} = \frac{1} {2} (3 w_d + 1)
\end{equation}
Thus, if $w_d > -1/3$ then the expansion of universe will decelerate; if $w_d < -1/3$, then the universe will eventually accelerate; if $w_d = -1/3$ then the universe will approach a coasting universe.

The second key result is the asymptotic behaviour of the integral for $\chi$ in Equation \eqref{eq:georad2} when $R(t)$ given by Equation \eqref{eq:domat}. Whilst the exact indefinite integral unfortunately involves the hypergeometric function, we can approximate this function in the limit of large $t$ as by noting that $\sqrt{1 + CR^2} \approx \sqrt{C}R$ in this limit. The integral then becomes trivial. We now analyse the seven definitions, in order from the weakest to the strongest conditions.

\textbf{Definition 1 ($\dot{\chi}\to 0$): } From Equation \eqref{eq:georad2} we can see that as $t \to \infty$ (and $R(t) \to \infty$), $\dot{\chi} \propto R^{-2}$. Thus Definition 1 holds in all eternally expanding universes, so that the velocity of a particle through coordinate space will always decay to zero. \citet{2006astro.ph..3162G} use this definition when they claim that test particles will rejoin the Hubble flow.

\textbf{Definition 3 ($v_{\textrm{pec}} \to 0$): } Since $v_{\textrm{pec}}(t) = R(t)\dot{\chi}(t)$, Definition 3 is stronger than Definition 1. However, it still holds in all eternally expanding universes, since Equation \eqref{eq:georad2} shows that as $t \to \infty$, $v_{\textrm{pec}}(t) \propto R^{-1}$. This is closely related to the well known result that the momentum of any particle in the universe decays as $R^{-1}$. It is this definition that is most widely used to justify the claim that test particles rejoin the Hubble flow asymptotically.

\textbf{Definition 7} (CMB dipole $ \to 0$):  Equation \eqref{eq:CMBtemp} shows that this definition is equivalent to Definition 3 and thus holds in all eternally expanding universes. In particular, this shows that the dipole in the CMB will decay faster than the CMB temperature itself.

\textbf{Definition 2 ($\chi \to \chi_{\infty}$): } At first sight, Definition 2 may appear to be a direct consequence of Definition 1---surely if the derivative of a function approaches zero then the function itself will approach a constant. However, it is easy to think of counterexamples: $f(x) = \log(x)$. Thus, we need to consider Equation \eqref{eq:georad2}, integrating between $t_0 = 1$ and $t$ and using the two approximations discussed at the start of this section. This gives:
\begin{align}
\chi(t) =& \chi_0 \pm \int_{1}^{t} \frac{\textrm{d}t} {R\sqrt{1 + CR^2}} \\
\approx& K_1 \pm \frac{1} {\sqrt{C}R_0^2} \int^{t} \frac{\textrm{d}t} {t^{2n}} \label{eq:approxchi1}
\end{align}
where $n = \frac{2} {3(1+w_d)}$ and $K_1$ is a constant that includes $\chi_0$ and the primitive of the exact integral evaluated at $t_0$. Thus, for the integral to be bounded as $t \to \infty$, we require that:
\begin{equation}
2n > 1 \quad \Rightarrow \quad w_d < 1/3
\end{equation}
Thus if we use Definition 2 to define what it means to asymptotically rejoin the Hubble flow, then test particles in universes where the dominant energy component has equation of state $w_d \ge 1/3$ do not rejoin the Hubble flow. In particular, in a universe where the dominant energy component is radiation, the comoving coordinate of a test particle removed from the Hubble flow increases (or decreases) without bound. The particle has no rightful place in the Hubble flow\footnote{\citet{2003gieg.book.....H} sets the derivation of Equation \eqref{eq:georad2} as a practice problem for undergraduates. The solutions give a very instructive derivation using Killing vectors, but then state that the particle comes to rest at coordinate $x_f \equiv \chi_{\infty}$, given by integrating from zero to infinity. He fails to note that the integral may diverge, so that the particle may not come to rest at all.}.

\textbf{Definition 4} ($\dot{r}_\textrm{p} \to v_{\textrm{rec}}$): Definition 4 appears to be identical to Definition 3; surely if $\dot{r}_\textrm{p}(t) = v_{\textrm{rec}}(t) + v_{\textrm{pec}}(t)$ and $v_{\textrm{pec}}(t) \to 0$ in all eternally expanding universes then Definition 4 holds trivially. However, it is possible that $v_{\textrm{rec}}(t)$ also goes to zero, and if it does so as fast or faster than $v_{\textrm{pec}}(t)$ then we are not justified in saying that the proper velocity of the test particle approaches its recession velocity. We can see this from Equation \eqref{eq:asympequiv}:
\begin{align}
\dot{r}_\textrm{p}(t) \to v_{\textrm{rec}}(t) \quad &\Rightarrow \quad \frac{\dot{r}_\textrm{p}(t)} {v_{\textrm{rec}}(t)} \to 1 \\
&\Rightarrow \quad \frac{v_{\textrm{rec}}(t) + v_{\textrm{pec}}(t)} {v_{\textrm{rec}}(t)} \to 1 \\
&\Rightarrow \quad \frac{ v_{\textrm{pec}}(t)} {v_{\textrm{rec}}(t)} \to 0
\end{align}
which is stronger than the condition in Definition 3. \citet[][page 11]{2004Obs...124..174W} hinted at Definition 4: ``Peculiar velocities do vanish eventually in expanding universes; but so do \emph{all} velocities [emphasis original]'', but mistakenly implied that it would rule out rejoining the Hubble flow in matter dominated universes ($w_d = 0$). \citet{2001astro.ph..4349D} use Definition 4 (see their equation (11) and following) but mistakenly believe that it follows automatically from the success of Definition 3.

As we noted previously, as $t \to \infty$, $v_{\textrm{pec}}(t) \propto R^{-1}$. Thus:
\begin{equation} \label{eq:vpecn}
v_{\textrm{pec}}(t) \propto R^{-1} \propto t^{-n}
\end{equation}
We now consider $v_{\textrm{rec}}(t) = \dot{R}(t)\chi(t)$. Using Equation \eqref{eq:approxchi1} we approximate $\chi$ by:
\begin{align}
& \chi(t) \approx K_1 \pm  \frac{1} {\sqrt{C}R_0^2}\begin{cases}
\frac{t^{1-2n}} {1-2n} & \text{if $n \neq 1/2$}\\
\log t & \text{if $n = 1/2$}
\end{cases} \label{eq:approxchi2}\\
\Rightarrow \quad &v_{\textrm{rec}}(t) \propto K_2 t^{n-1} + K_3 \begin{cases}
t^{-n}& \text{if $n \neq 1/2$}\\
t^{-1/2}\log t & \text{if $n = 1/2$} \label{eq:approxvrec}
\end{cases}
\end{align}
where the $K_i$ will be used keep track of constants. Now, we need to consider three cases:
\begin{itemize}
\item If $n > 1/2$, then $n-1 > -n$ . Thus the dominant term in Equation \eqref{eq:approxvrec} is the first term so that $v_{\textrm{rec}}(t) \propto t^{n-1}$ for large $t$. Then, as $t \to \infty$:
\begin{equation}
\frac{v_{\textrm{pec}}(t)} {v_{\textrm{rec}}(t)} \propto \frac{t^{-n}} {t^{n-1}} \to 0
\end{equation}
Thus when $n > 1/2$, $v_{\textrm{pec}}(t)$ goes to zero faster than $v_{\textrm{rec}}(t)$, meaning that Definition 4 holds.
\item If $n = 1/2$, then $v_{\textrm{rec}}(t) \propto t^{-1/2}\log t$. Then, as $t \to \infty$:
\begin{equation}
\frac{v_{\textrm{pec}}(t)} {v_{\textrm{rec}}(t)} \propto \frac{t^{-1/2}} {t^{-1/2}\log t} \to 0
\end{equation}
Thus Definition 4 holds when $n = 1/2$.
\item If $n < 1/2$, then $n-1 < -n$ so that $v_{\textrm{rec}}(t) \propto t^{-n}$. Then, as $t \to \infty$:
\begin{equation}
\frac{v_{\textrm{pec}}(t)} {v_{\textrm{rec}}(t)} \propto \frac{t^{-n}} {t^{-n}} \to 1
\end{equation}
Thus when $n < 1/2$, $v_{\textrm{rec}}(t)$ and $v_{\textrm{rec}}(t)$ approach zero at the same rate. It is not true that $\dot{r}_\textrm{p}(t) \to v_{\textrm{rec}}(t)$ as $t \to \infty$ and Definition 4 fails in this case.
\end{itemize}

If we use Definition 4 to define what it means to asymptotically rejoin the Hubble flow, then test particles in universes where the dominant energy component has equation of state $w_d > 1/3$ do not rejoin the Hubble flow. Note that Definition 4 holds in a universe where the dominant energy component is radiation, unlike Definition 2, which fails.

\begin{figure*}
\centering
\begin{minipage}[c]{0.48\textwidth}
\centering
\includegraphics[width=\textwidth]{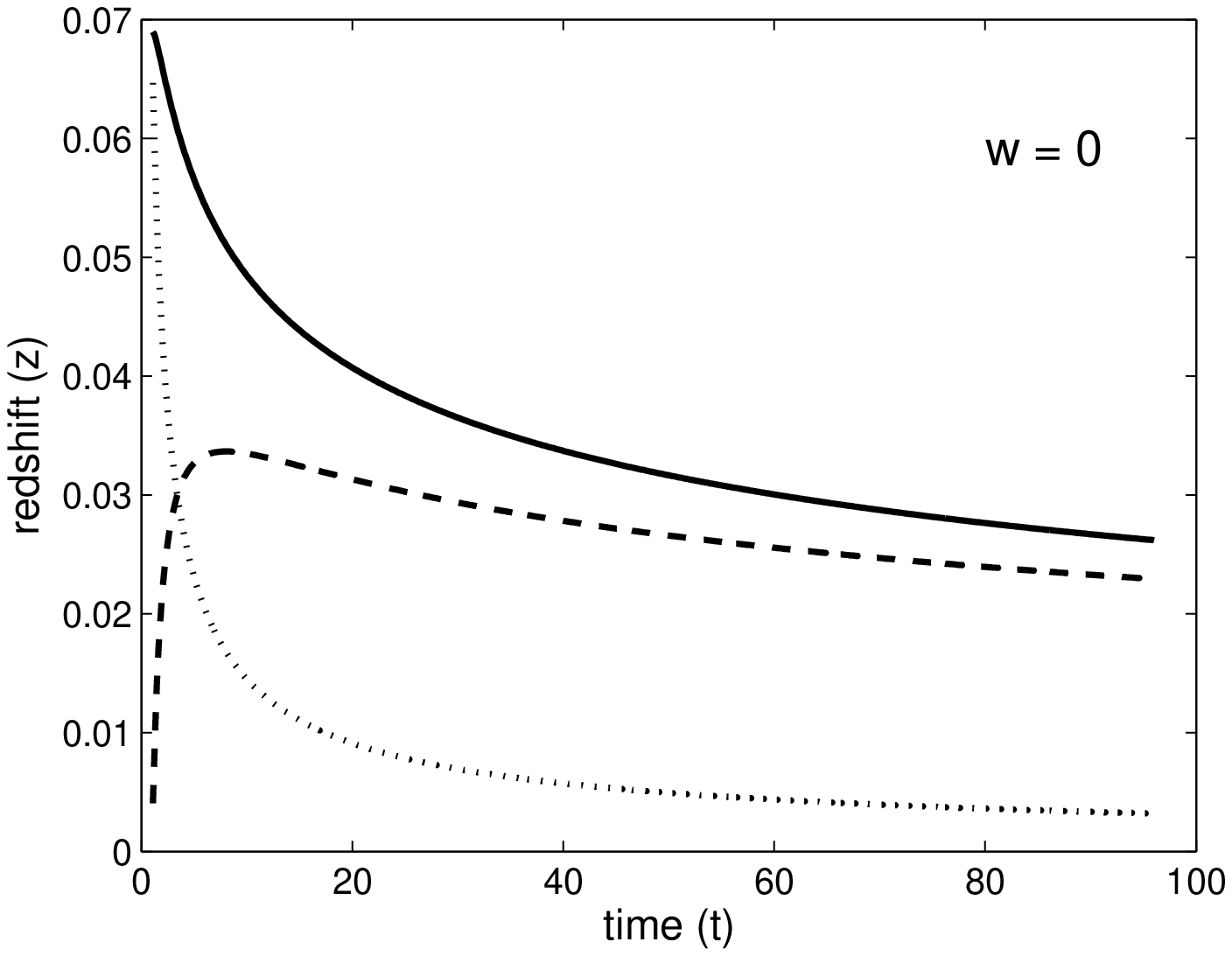}
\end{minipage}
\hspace{0.2cm}
\begin{minipage}[c]{0.48\textwidth}
\centering
\includegraphics[width=\textwidth]{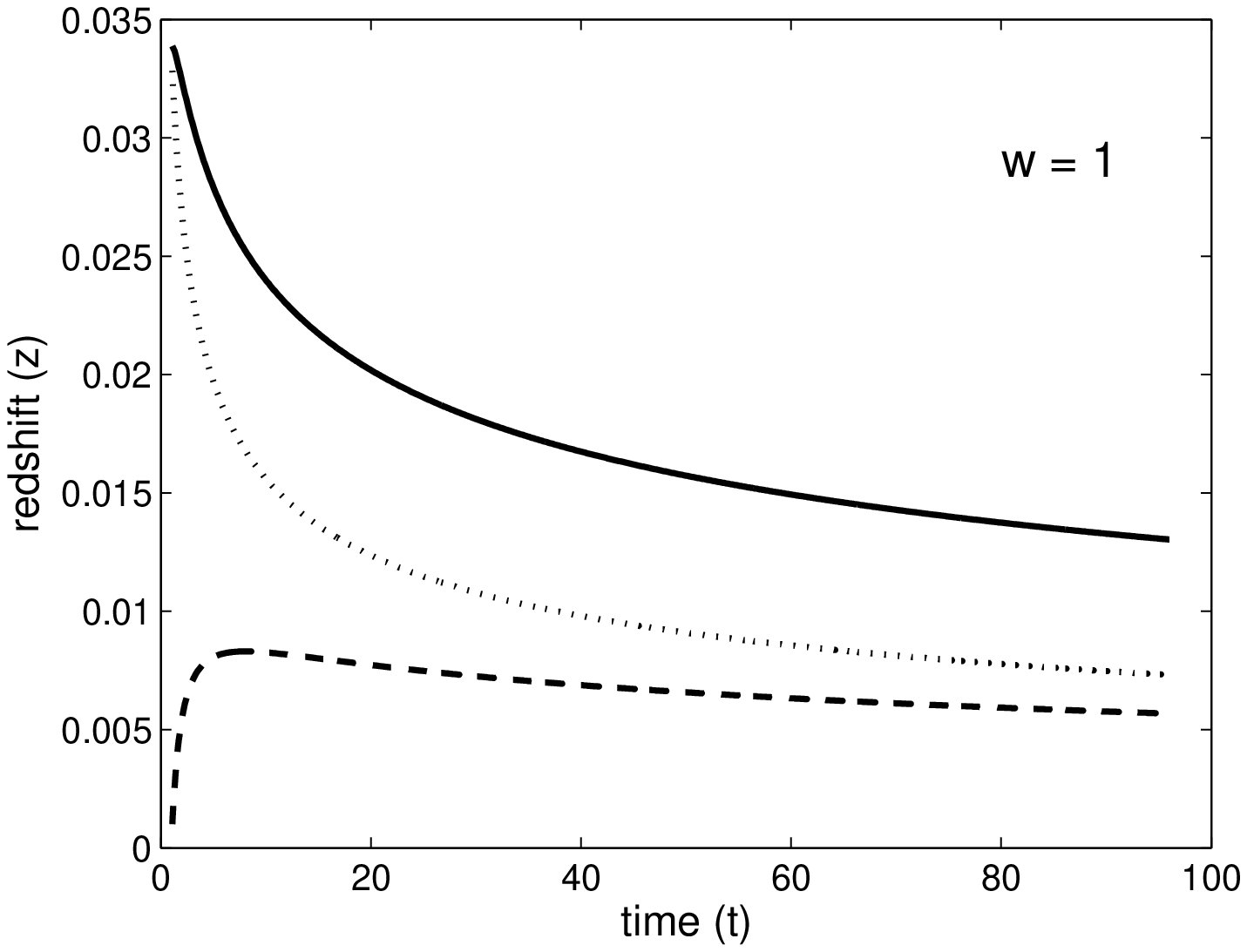}
\end{minipage}
\caption{Redshift as a function of time for cosmological models with $w=0$ (left) and $w=1$ (right) (examples of fluids with $w = 1$ are free massless scalar fields and shear energy, such as superhorizon gravitational waves; see \citet{Linder:1997}). The dashed line is the cosmological redshift ($z_{\textrm{cosm}}$), the dotted line is the Doppler redshift ($z_{\textrm{Dop}}$) and the solid line is the observed redshift ($z_{\textrm{obs}} = -1 + (1+z_{\textrm{cosm}})(1+z_{\textrm{Dop}})$). The left panel shows the asymptotic dominance of the cosmological redshift for $w \le 1/3$, whilst the right panel shows the identical asymptotic behaviour of the cosmological and Doppler redshift at large $t$ for $w > 1/3$.  }\label{fig:doppler}
\end{figure*}

\textbf{Definition 6} ($z_{\textrm{obs}} \to z_{\textrm{cosm}}$): Once again we are tempted to assume that the success of Definition 3 will ensure that this definition will hold in all universes. The argument proceeds as before: if
\begin{equation}
1+z_{\textrm{Dop}} = \left( \frac{1+v_{\textrm{pec}}(t_e)} {1-v_{\textrm{pec}}(t_e)} \right)^{\frac{1} {2}}
\end{equation}
and $v_{\textrm{pec}}(t) \to 0$ in all eternally expanding universes then Definition 6 holds trivially. And our \emph{caveat} is the same---we must be careful of the case where $z_{\textrm{cosm}}$ also goes to zero.

To do this, we need to express $z_{\textrm{obs}}$ in terms of reception time $t_r$ and then consider the limit $t_r \to \infty$. In fact it is much easier to express everything in terms of the emission time $t_e$ and then consider $t_e \to \infty$; $t_e < t_r$ guarantees that both cases will have identical limiting behaviour.

In the limit of small $v_{\textrm{pec}}$, we have that $z_{\textrm{Dop}} \sim v_{\textrm{pec}} \propto t^{-n}$, using Equation \eqref{eq:vpecn}. Also, $1+z_{\textrm{cosm}} = \frac{R(t_r)} {R(t_e)} = \frac{t_r^{n}} {t_e^{n}}$, so that the task at hand is to express $t_r$ in terms of $t_e$. Consider a light ray travelling along a null ($\textrm{d}s = 0$), ingoing ($\textrm{d}\chi < 0$), radial ($\textrm{d}\theta = \textrm{d}\phi = 0$) geodesic. From the RW metric:
\begin{align}
\textrm{d}t =& -R(t)\textrm{d}\chi  \\
\Rightarrow \chi(t_e) - \chi(t_r) =& \int^{t_r}_{t_e}\frac{\textrm{d}t} {R(t)}
\end{align}
where we now place the receiver at the origin: $\chi(t_r) = 0$. For the case of a test particle moved removed from the Hubble flow, $\chi(t_e)$ is given by Equation \eqref{eq:georad2}, approximated by Equation \eqref{eq:approxchi2}. An immediate consequence of combining these equations is that Definition 6 must work in accelerating and coasting universes ($w_d \le -1/3$), since in these universes $z_{\textrm{cosm}}$ does not go to zero. Thus we need only consider $0<n<1$. We will leave the $n = 1/2$ case to the reader: it involves $\log t$ as with Definition 4.

With this in mind, we can derive an expression for $t_r$:
\begin{equation}
t_r = \left(t_e^{1-n} + K_4 t_e^{1-2n}+ K_5\right)^{\frac{1} {1-n}}
\end{equation}
which leads to the following expression for $1 + z_{\textrm{cosm}}$:
\begin{align}
1 + z_{\textrm{cosm}} &= \left(1 + K_4 t_e^{-n}+ K_5 t_e^{-(1-n)}\right)^{\frac{n} {1-n}}\\
\approx& 1 + K_6 t_e^{-n}+ K_7 t_e^{-(1-n)}
\end{align}
where the last expression is calculated using the binomial theorem for non-integer exponents\footnote{See \citet{mathworld} for a reminder.}. This leads to the following three cases:
\begin{itemize}
\item If $n > 1/2$, then $n-1 > -n$ so that $z_{\textrm{cosm}} \propto t^{n-1}$. Then, as $t \to \infty$:
\begin{equation}
\frac{z_{\textrm{Dop}}} {z_{\textrm{cosm}}} \propto \frac{t^{-n}} {t^{n-1}} \to 0
\end{equation}
Thus when $n > 1/2$, $z_{\textrm{Dop}}$ goes to zero faster than $z_{\textrm{cosm}}$, meaning that Definition 6 holds.
\item If $n = 1/2$, it turns out that $z_{\textrm{cosm}} \propto t^{-1/2}\log t$. Then, as $t \to \infty$:
\begin{equation}
\frac{z_{\textrm{Dop}}} {z_{\textrm{cosm}}} \propto \frac{t^{-1/2}} {t^{-1/2}\log t} \to 0
\end{equation}
Thus Definition 4 holds when $n = 1/2$.
\item If $n < 1/2$, then $n-1 < -n$ so that $z_{\textrm{cosm}} \propto t^{-n}$. Then, as $t \to \infty$:
\begin{equation}
\frac{z_{\textrm{Dop}}} {z_{\textrm{cosm}}} \propto \frac{t^{-n}} {t^{-n}} \to 1
\end{equation}
Thus when $n < 1/2$, $z_{\textrm{cosm}}$ and $z_{\textrm{Dop}}$ approach zero at the same rate. It is not true that $z_{\textrm{obs}} \to z_{\textrm{cosm}}$ and Definition 6 fails in this case.
\end{itemize}

If we use Definition 6 to define what it means to asymptotically rejoin the Hubble flow, then test particles in universes where the dominant energy component has equation of state $w_d > 1/3$ do not rejoin the Hubble flow. The situation is illustrated in Figure \ref{fig:doppler}. When $w=0$ the Doppler redshift decays away much faster than the cosmological redshift, so that Definition 6 holds. However, when $w=1$ the Doppler and the cosmological redshift decay at the same rate; which one is greater depends on the cosmological model and the initial conditions (hidden in the constants $K_i$). Thus, Definition 6 fails in this universe.

\textbf{Definition 5} ($\Delta r_{\textrm{p}}\to 0$): We know already that this definition fails in some cases since it relies on Definition 2. Thus we begin with the assumption that $w_d < 1/3$, i.e. $n > 1/2$. Now, we have that:
\begin{align}
\Delta r_{\textrm{p}} =& \lvert R(t)\chi_{\infty} - R(t)\chi(t) \rvert \\
=& R(t) \bigg\lvert \int_{t_0}^{\infty} \frac{\textrm{d}t} {R\sqrt{1 + CR^2}} - \int_{t_0}^t \frac{\textrm{d}t}  {R\sqrt{1 + CR^2}} \bigg\rvert \\
=& R(t) \bigg\lvert \int_{t}^{\infty} \frac{\textrm{d}t} {R\sqrt{1 + CR^2}} \bigg\rvert \\
\propto& \ t^n \bigg\lvert \int_{t}^{\infty} \frac{\textrm{d}t} {t^{2n}} \bigg\rvert \\
\propto& \ t^n \ t^{1-2n} = t^{1-n}
\end{align}

Hence the requirement that $\Delta r_{\textrm{p}} \to 0$ as $t \to \infty$ is only met for $n > 1$, i.e. $w_d < -1/3$ or $q < 0$. In particular, for a universe where the dominant component has $w_d = -1/3$, $\Delta r_{\textrm{p}}$ approaches a constant. For example, \citet{2004Obs...124..174W}, page 10 reaches this conclusion for an underdense, matter only universe.  If we use Definition 5 to define what it means to asymptotically rejoin the Hubble flow, then test particles in universes where the dominant energy component has equation of state $w_d \ge -1/3$ do not rejoin the Hubble flow. This means that particles asymptotically rejoin the Hubble flow only in universes that eventually accelerate. Definition 5 is the strongest of the definitions, and is the one used by \citet{2004Obs...124..174W}, page 10, to justify the claim that particles in a matter dominated universe ($w_d =0$) do not join the Hubble flow.

The failure of this definition can be illustrated by looking again at Figure \ref{fig:chi}. The first column ($w = -2/3$) clearly shows a particle that satisfies Definition 5. However, in the second and third columns ($w=-1/3,0$), in the bottom panels, the lowest dashed line is the trajectory of the particle in the Hubble flow whose comoving coordinate is equal to $\chi_{\infty}$, i.e. the rightful place of the particle in the Hubble flow. In the centre column, the test particle trajectory and the rightful place trajectory move apart, whilst in the second column they are separated by a constant. Thus, the particle never joins the Hubble flow in the sense of its trajectory being indistinguishable from the trajectory of a particle in the Hubble flow.

A summary of the different definitions and the conditions for their fulfilment is given in Table \ref{deftable}.
\begin{table}
\centering
\caption{Seven definitions and the conditions for their fulfilment.}\label{deftable}
\begin{tabular}{|c|c|c|c|}
\hline
 & 				 & Hold in all				 & If no, conditions \\
 & Definition	 &  expanding 					& on $w_d$ for \\
 &             & universes?       		 & definition to hold \\
\hline
1& $\dot{\chi}\to 0$ & yes & \\
\hline
2& $\chi \to \chi_{\infty}$ & no & $w_d <1/3$\\
\hline
3& $v_{\textrm{pec}} \to 0$ &  yes & \\
\hline
4& $\dot{r}_\textrm{p} \to v_{\textrm{rec}}$ &  no & $w_d \le 1/3$\\
\hline
5& $\Delta r_{\textrm{p}}\to 0$ &  no & $w_d < -1/3$\\
\hline
6& $z_{\textrm{obs}} \to z_{\textrm{cosm}}$ &  no & $w_d \le 1/3$\\
\hline
7& CMB dipole $ \to 0$ &  yes & \\
\hline
\end{tabular}
\end{table}


\section[space]{Particle Motion, the Hubble Flow and Expanding Space}\label{space}
The attack on the physical concept of expanding space has centred on the motion of test particles in the universe, as discussed in Section \ref{testparticles}. \citet{2004Obs...124..174W}, \citet{Peacockweb}  and others claim expanding space fails to adequately explain test particle motion because we expect that expanding space will carry the particle away from the origin, as stretching rubber would carry the glowing ball of Section \ref{testparticles} away. Whilst we will not attempt to resolve this debate here, we believe that this may be a misunderstanding of expanding space that is fostered by a flaw in the rubber sheet analogy. Particles in the Hubble flow do not feel any force, as they are free-falling. Thus, thinking of expanding space as a frictional or viscous force (like objects on a rubber sheet) is incorrect.

Finally, how are we to understand the failure of so many of the definitions of joining the Hubble flow in Section \eqref{joininghubble}? We contend that expanding space predicts that the peculiar velocity of a test particle will approach zero \citep{Peacockweb}:
\begin{quote}
through the accumulated Lorentz transforms required to overtake successively more distant particles that are moving with the Hubble flow.
\end{quote}
Thus we expect Definition 3 (along with Definitions 3 and 7, which are weaker or equivalent conditions) to hold in all eternally expanding universes. However, expanding space does not lead us to expect that Definitions 2, 4, 5 and 6 will hold. We contend that the problem is not that expanding space has mislead us, but that describing the decay of $v_{\textrm{pec}}$ as joining the Hubble flow is a misnomer. The correspondence between this particular characteristic of the trajectory of the test particle and the trajectory of particles in the Hubble flow leads us to expect that all of the features of the trajectory ($\chi$, $\dot{\chi}$, $r_\textrm{p}$, $\dot{r}_\textrm{p}$, $v_{\textrm{pec}}$) will approach those of the Hubble flow trajectories (respectively, $\chi=\chi_{\infty}$, $\dot{\chi}=0$, $r_\textrm{p}=R(t)\chi_{\infty}$, $\dot{r}_\textrm{p}=v_{\textrm{rec}}$, $v_{\textrm{pec}}=0$). But this is too much to ask from the expansion of the universe. As we have seen, many of these conditions depend on the acceleration and deceleration of the universe, rather than just its expansion.

We contend that the correct definition of asymptotically rejoining the Hubble flow is that all the features of the test particle trajectory approach the corresponding features of the Hubble flow trajectories. Selecting one feature of the trajectory on which to base our definition is arbitrary and leads to a multitude of conflicting claims. All seven definitions have equal claim to the title of \emph{the} definition of joining the Hubble flow. We therefore require that all definitions hold, which is equivalent to just requiring Definition 5 to hold, as it is the strongest definition. It follows that it is not a general feature of expanding universes that test particles asymptotically rejoin the Hubble flow.

\section*{Acknowledgments}
LAB and JBJ are supported by an Australian Postgraduate Award. LAB is also supported by a University of Sydney School of Physics Denison Merit Award. JBJ is also generously supported by St Andrew's College within the University of Sydney. MJF is supported by a University of Sydney Faculty of Science UPA scholarship. GFL gratefully acknowledges support through the Institute of Astronomy visitor program.

\bsp

\label{lastpage}

\end{document}